\documentclass[12pt]{article}
\usepackage[dvips]{graphicx}
\usepackage{latexsym}
\usepackage[abs]{overpic}
\usepackage{wrapfig}
\usepackage{amssymb}
\usepackage{amsmath}

\input epsf
\newcommand{\R}{R}
\newcommand{\sect}[1]{\section{#1}\setcounter{equation}{0}}

\makeatletter
\@addtoreset{equation}{section} 
\makeatother

\newcommand{\skyp}[1]{}

\def\Z {\bb{Z}}
\def\R {\bb{R}}

\def\rem#1{}
\def\C{{\mathbb{C}}} % \C=\mathbb{C}
\def\R{{\mathbb{R}}} % \R=\mathbb{R}
\def\Z{{\mathbb{Z}}} % \Z=\mathbb{Z}
\def\D{{\mathcal{D}}} % \D=\mathcal{D} 

\def\lam{{\lambda}}
\def\Tr{{\mathrm{Tr}}}
\def\del{{\partial}}

\newcommand\non{\nonumber \\}
\newcommand{\bel}{\begin{eqnarray}}
\newcommand{\ee}{\end{eqnarray}}

\def\del{{\partial}} %  

\begin{document}

\begin{titlepage}

\bigskip
\hfill\vbox{\baselineskip12pt
\hbox{KEK-TH-1569}
\hbox{RUP-12-10}
}
\bigskip\bigskip\bigskip\bigskip

\begin{center}
{\LARGE \bf $G/G$ gauged WZW model and Bethe Ansatz for the phase model}
\end{center}

\bigskip
\bigskip

\bigskip
\bigskip
\centerline{ \large Satoshi Okuda$^1$ and Yutaka Yoshida,$^2$}
\bigskip
\medskip
\centerline{$^1$\it Department of Physics, Rikkyo University}
\centerline{\it Toshima, Tokyo 171-8501, Japan, \ \ }
\centerline{ okudas@rikkyo.ac.jp}
\bigskip
\centerline{$^2$\it High Energy Accelerator Research Organization (KEK)}
\centerline{\it Tsukuba, Ibaraki 305-0801, Japan}
\centerline{ yyoshida@post.kek.jp}
\bigskip
\bigskip

\bigskip
\bigskip

\begin{abstract}
We investigate the G/G gauged Wess-Zumino-Witten model on a Riemann surface
from the point of view of the algebraic Bethe Ansatz for the phase model.
After localization procedure is applied to the G/G gauged Wess-Zumino-Witten model, the diagonal components
for group elements satisfy Bethe Ansatz equations for the phase model. 
We show that the partition function of the G/G gauged Wess-Zumino-Witten model is identified as the summation of norms with respect to all the eigenstates of the Hamiltonian with the fixed number of particles in the phase model.
We also consider relations between the Chern-Simons theory on $S^1\times\Sigma_h$ and the phase model.
\end{abstract}

\end{titlepage}

\newpage
\baselineskip=18pt

\sect{Introduction}

It is known that there exist connections between topological gauge theories and the Bethe Ansatz  for integrable systems (Gauge/Bethe correspondence).
In \cite{Moore:1997dj}, the equivariant localization for the BF theory coupled to a one-form valued  adjoint Higgs field (Yang-Mills-Higgs system) is studied 
and the localization configurations lead to Bethe Ansatz equations for the non-linear Schr$\ddot{\text{o}}$dinger model. 
Further, the partition function of the Yang-Mills-Higgs system is
related to norms of wave functions for the non-linear Schr$\ddot{\text{o}}$dinger model 
\cite{Gerasimov:2006zt}. 
So we expect that the partition functions of other topological gauge theories  are also related to norms of wavefunctions for the corresponding integrable systems.

Wess-Zumino-Witten (WZW) model of a two-dimensional conformal field theory has rich structures and various applications in mathematics and also in physics.
For example, the Hilbert space of the Chern-Simons (CS) theory with a gauge group $G$ on $\R \times \Sigma_h$ is equivalent to the space of the conformal block for the $G/G$ gauged WZW model on a Riemann surface $\Sigma_h$.
The partition function of the CS theory on a three manifold  can be  obtained by  sewing  the  boundary Riemann surfaces which is
implemented by an inner product of states on $\Sigma$.     
One can also calculate Wilson loop expectation  values which give knot invariants in terms of fusion coefficients and modular matrices \cite{Witten:1988hf}.

In the WZW model, one can construct the $G/H$ gauged WZW model by gauging an anomaly free subgroup $H$ of the global symmetry group $G$.
The $G/H$ gauged WZW model is an explicit lagrangian realization of the coset construction in the CFT.
When $H=G$, the $G/G$ gauged WZW model becomes a topological field theory \cite{Witten:1991mm}, \cite{Spiegelglas:1992jg}.
One  expects  that there exists a  method for calculating the partition function and correlation functions without relying on the CFT techniques nor the representation theory of the affine Lie algebra. 
Actually, these were derived by a field theoretic approach in \cite{Blau:1993tv}, \cite{Gerasimov:1993ws}. 
 In this approach,  it is important for the $G/G$ gauged WZW model to possess  a certain BRST-type symmetry whose square generates a $G$-gauge transformation. 
 This symmetry makes it possible to work out the path integrals with insertions of BRST closed operators via equivariant localization procedure. 
In higher rank of the gauge group, the localization configurations for the diagonal components of $G$-elements   are complicated. However, the final expression for the partition function is  simply expressed by  modular matrices and  an integrable structure  might inhibit behind the localization configurations.
 
In this paper, we will show that the integrable system corresponding to the $U(N)/U(N)$ gauged WZW model is 
the phase model.
The phase model is an integrable field theory on one-dimensional lattice.
We can apply the algebraic Bethe Ansatz.
For example, see \cite{Bogoliubov:1996}, \cite{Bogoliubov:1997}.
It is known that the phase model appears in the $SU(N)$ WZW model.
Recently,  Korff and Stroppel established the $\widehat{\mathfrak{su}}(N)_k$ Verlinde algebra \cite{Verlinde:1988sn} in terms of the algebraic Bethe Ansatz for the phase model and derived an efficient  recursion relation for calculating  fusion coefficients \cite{Korff:2010}. 
See also a short review \cite{Walton:2012pi}.
We will consider relations between the Gauge/Bethe correspondence and \cite{Korff:2010}.
We also point out that the partition function of the CS theory on $S^1\times \Sigma_h$ is  related to norms of  Hamiltonian eigenstates for the phase model.

This paper is organized as follows. 
In section \ref{sec:GWZW}, we study the localization for the $U(N)/U(N)$ gauged WZW model and
 find that  the localization configurations satisfy particular equations. 
In section \ref{sec:phase}, we review the algebraic Bethe Ansatz for the phase model.
In section \ref{sec:GWZW-phase}, 
we find that the Bethe Ansatz equations equal to the localization configurations of the $U(N)/U(N)$ gauged WZW model under parameter identifications. 
We find relations between the partition function of the $U(N)/U(N)$ gauged WZW model and the Bethe norms for the phase model.
The section \ref{sec:summary} is devoted to conclusion.

\section{$G/G$ gauged Wess-Zumino-Witten model and localization}
\label{sec:GWZW}

In this section, we calculate the partition function 
of the $G/G$ gauged WZW model on a genus-$h$ Riemann surface  $\Sigma_h$ by applying localization method.
This model is a topological field theory and 
its partition function counts the number of conformal blocks in the $G$ WZW model with level $k$ on $\Sigma_h$.
When we set $g=e^{i \phi/k}$ and expand them at the leading order of $1/k$,
the $G/G$ gauged WZW model on $\Sigma_h$ reduces to the BF theory with the gauge group $G$ on $\Sigma_h$ \cite{Witten:1991we}.
This can be regarded as an another deformation of the BF theory different from the Yang-Mills-Higgs system.

The partition function of the $G/G$ gauged WZW model on $\Sigma_h$ is defined by  
\begin{eqnarray}
Z_{\rm GWZW}^G(\Sigma_h)=\int \mathcal{D}g \mathcal{D}A \mathcal{D} \lambda  e^{-k S_{G/G}(g,A,\lambda)}.
\label{partitionGWZW}
\end{eqnarray}
The action of the $G/G$ gauged WZW model is 
\begin{eqnarray}
 S_{{\rm GWZW}}(g,A,\lam)&=&S_{\rm WZW}(g) 
 -\frac{1}{2\pi} \int_{\Sigma_h} d^2z 
 \Tr (A_z  \del_{\bar{z}} g g^{-1}  -  A_{\bar{z}} g^{-1} {\del}_{{z}} g
 -g^{-1} A_z g A_{\bar{z}} +A_{z} A_{\bar{z}}) \nonumber\\
&& - \frac{1}{2\pi}\int_{\Sigma_h}d^2 z \Tr(\lambda_z\lambda_{\bar{z}})
\end{eqnarray}
with
\begin{eqnarray}
S_{\rm WZW} (g)&=&-\frac{1}{4\pi} \int_{\Sigma_h} d^2z  \Tr (g^{-1}  {\del}_{{z}} g \cdot  g^{-1}  {\del}_{\bar{z}} g ) -i  \Gamma(g)
\end{eqnarray}
and the Wess-Zumino term $\Gamma(g)$ is
\begin{eqnarray}
\Gamma(g) = \frac{1}{12\pi} \int_B d^3y \epsilon^{i j k} \Tr\left( g^{-1} \del_i g \cdot g^{-1} \del_j g  \cdot g^{-1} \del_k g \right).
\end{eqnarray}
Here $B$ is a three dimensional manifold with the boundary $\partial B=\Sigma_h$.
$g(z,\bar{z})$ is the $G$-valued field, $A=A_z dz +A_{\bar{z}} dz$ is a two-dimensional gauge field and $\lambda=\lambda_{z} dz + \lambda_{\bar{z}}d\bar{z}$
is a one-form adjoint fermion.  We also denote the holomorphic part of $A$ as $A^{(1,0)}=A_z dz$ and anti-holomorphic part as 
$A^{(0,1)}=A_{\bar{z}} dz$ and so on. $Q$ is a scalar BRST charge defined as
\begin{eqnarray}
Q A = \lambda, \quad Q\lambda^{(1,0)} = (A^g)^{(1,0)}- A^{(1,0)}, 
\quad Q\lambda^{(0,1)} =- (A^{g^{-1}})^{(0,1)}+  A^{(0,1)}, \quad Q g=0
\nonumber\\
\end{eqnarray}
with $A^g=g^{-1} d g+ g A g^{-1}$. 
The partition function (\ref{partitionGWZW}) is invariant under the BRST transformation.
The square of the BRST transformation generates following gauge transformations
\begin{eqnarray}
&&{\cal L}_g A^{(1,0)} = (A^g)^{(1,0)} - A ^{(1,0)},
\quad {\cal L}_g A^{(0,1)} =  -(A^{g^{-1}})^{(0,1)} + A ^{(1,0)},\nonumber\\
&&{\cal L}_g \lambda^{(1,0)} = g^{-1}\lambda^{(1,0)}g - \lambda^{(1,0)},
\quad {\cal L}_g \lambda^{(0,1)} = -g\lambda^{(0,1)}g^{-1} + \lambda^{(0,1)},
\quad {\cal L}_g g =0,
\end{eqnarray}
where $Q^2 = {\cal L}_g$.

From now on, we set the gauge group $G$ as $U(N)$ for simplicity.
We evaluate the partition function (\ref{partitionGWZW}).  
The resulting expression is
\begin{eqnarray}
Z_{\rm GWZW}^{U(N)}(\Sigma_h) &=&
\frac{1}{N!} \sum_{n_1,\cdots,n_N =-\infty}^\infty\int \prod_{a=1}^{N}d \phi_a \prod_{\substack{a,b=1\\a \neq b}}^N 
(1-e^{2\pi i (\phi_a -\phi_b)})^{1-h} \times \nonumber\\
&&\hspace{1.5cm}
\times \exp\left\{2\pi i \sum_{a=1}^N n_a 
    \left((N+k)\phi_a - \sum_{b=1}^N\phi_b + \frac{N-1}{2}
    \right)
\right\}.
 \label{contourpartition}
\end{eqnarray}

Let us  briefly explain the derivation of (\ref{contourpartition}).
See  detailed calculation \cite{Blau:1993tv}.
First of all, we take a diagonal gauge on $g(z,\bar{z}) \in U(N)$ at (\ref{partitionGWZW}), that is, 
$g(z,\bar{z})=e^{2\pi i\sum_{a=1}^N \phi_a H^a}$ where $H^a$ are the Cartan generators. 
Then the path integral with respect to $g(z,\bar{z})$ becomes
\begin{eqnarray}
\int \mathcal{D} g f(g) =\frac{1}{N!}\int \prod_{a=1}^N \D \phi_a \mathrm{Det}_{c, \bar{c} } (1-\mathrm{ad}(e^{2\pi i \phi}))   f(e^{ 2\pi i \phi }),
\end{eqnarray}
where $|W| = N!$ is the order of Weyl group for the gauge group $U(N)$.
Here $\mathrm{Det}_{c, \bar{c} }$ is the Faddeev-Popov determinant for the diagonal gauge fixing.
Next, the path integral with respect to off-diagonal components of the gauge field gives a functional determinant  
\begin{eqnarray}
\mathrm{Det}^{-1}_{A^{(1,0)}, A^{(0,1)}} (1-\mathrm{ad}(e^{2\pi i \phi})).
\label{oneloopgauge}
\end{eqnarray}
The each functional determinant of the ghost and the gauge field are infinite dimensional and divergent, but
their differences can be explicitly evaluated by the Hirzebruch-Riemann-Roch theorem:
\begin{eqnarray}
\frac{\mathrm{Det}_{c, \bar{c}} ({1}-  \mathrm{ad} (e^{2\pi i  \phi}) ) }{ \mathrm{Det}_{A^{(1,0)}, A^{(0,1)} } (1- \mathrm{ad } (e^{2\pi i  \phi }))}
=\prod_{\substack{a,b=1\\a\neq b}}^N
(1- e^{2\pi i(\phi_a -\phi_b)})^{1-h+n_a-n_b}.
\end{eqnarray}
where  $n_a$ is  an $a$-th diagonal $U(1)$-charge of the background gauge field:
\begin{eqnarray}
n_a =\frac{1}{2 \pi}\int_{\Sigma_h} F^{(a)}.
\end{eqnarray}
Combining these results, the partition function of the $U(N)/U(N)$ gauged WZW model is 
\begin{eqnarray}
\int \D g \D A \D \lambda e^{- k S_{\rm GWZW}} &=&\frac{1}{N!}\sum_{(n_1,\cdots,n_N) \in \Z^N} 
\int \prod_{i =1}^N \D \phi_a \D A^{(a)}   \D \lambda
\prod_{\substack{a,b=1\\a\neq b}}^N \bigl( 1- e^{2 \pi i (\phi_a- \phi_b)}  \bigr)^{1-h+n_a-n_b}  \non
&& \quad \times \exp \Bigl(i k \sum_{a=1}^{N} \int_{\Sigma_h} \phi_a F^{a} - \frac{i k}{2\pi} \int_{\Sigma_h} \lambda \wedge \lambda \Bigr).
\label{effectiveaction} 
\end{eqnarray}
Thus, the effective abelianized action obtained the BF-type action
\begin{eqnarray}
i \sum_{a=1}^N \int_{\Sigma_h} \phi_a F^{a}. 
\end{eqnarray}

Here we note the fermion bilinear term $\lambda \wedge \lambda$.
Since the effective action (\ref{effectiveaction}) is not BRST invariant, 
we have to add appropriate counter terms  to restore the BRST symmetry 
by requiring the effective action to satisfy descent equations.
This leads to level-shift $k \to k+N$ for the coefficient of the fermion bilinear term. 
Of course, the such renormalization do not makes influences on the later calculation
because the fermion bilinear term enters in the effective action  freely.
The renormalization effect becomes crucial when we couple the theory with additional matters
\cite{Moore:1997dj} \cite{Miyake:2011yr}.

$F^b$ can be always decomposed to a
harmonic part $F^{(b)}$ and an exterior
derivative of a one-form $d a_b$ such that
\begin{eqnarray}
F^b=F^{(b)} +d a_b.
\end{eqnarray}
Integrating $a_b $ by part puts delta functional constraints on $d \phi_a$ 
, so the fields $\phi_a(z, \bar{z})$ reduce to constant fields. Therefore, we obtain (\ref{contourpartition}).

We rewrite (\ref{contourpartition}) by using the Poisson resummation formula:
\begin{eqnarray}
Z_{\rm GWZW}^{U(N)}(\Sigma_h) &=& \frac{1}{N!}
 \sum_{m_1,\cdots,m_N =-\infty}^\infty\int \prod_{a=1}^{N}d \phi_a \prod_{\substack{a,b=1\\a \neq b}}^N 
(1-e^{2\pi i (\phi_a -\phi_b)})^{1-h}\times\nonumber\\
&& \hspace{2cm}\times
 \prod_{a=1}^N \delta\left((N+k) \phi_a - \sum_{b=1}^N \phi_b + \frac{N-1}{2} -m_a\right).
\label{partition del}
\end{eqnarray}
The partition function (\ref{partition del}) is invariant under the interchange $k \leftrightarrow N$ 
because the $U(N)/U(N)$ gauged WZW model on $\Sigma_h$ has a property of the level-rank duality \cite{Naculich:2007nc}.
Therefore, we can rewrite (\ref{partition del}) as
\begin{eqnarray}
Z_{\rm GWZW}^{U(N)}(\Sigma_h) &=&
\frac{1}{k!}
 \sum_{m_1,\cdots,m_k =-\infty}^\infty\int \prod_{a=1}^{k}d \phi_a \prod_{\substack{a,b=1\\a \neq b}}^k 
(1-e^{2\pi i (\phi_a -\phi_b)})^{1-h}\times\nonumber\\
&&\hspace{2cm}\times
 \prod_{a=1}^k \delta\left((N+k) \phi_a - \sum_{b=1}^k \phi_b + \frac{k-1}{2}-m_a\right).
\label{partition del2}
\end{eqnarray}
Integrating (\ref{partition del2}) with respect to $\phi_a$s,
the partition function localizes to configurations which the constant fields $\phi_a$ satisfy constraints
\begin{eqnarray}
(N+k) \phi_a - \sum_{b=1}^k \phi_b + \frac{k-1}{2}-m_a=0.
\label{constraint}
\end{eqnarray}

Let us consider solutions of these equations.
We immediately find that the solutions
are
\begin{eqnarray}
\phi_a
=\frac{1}{k+N}\left(J_a + \frac{1}{N}||J||\right),
\label{Solution}
\end{eqnarray}
where $J_a = m_a - \frac{k-1}{2}$, $m_a \in \Z$ and $||J|| = \sum_{a=1}^k J_a$ for $a=1,\cdots,k$.
Note that range of the each field $\phi_a$ is $0 \le \phi_a <1$ .
The  fields in this range only contribute to the partition function.
Therefore, we count the number of solutions  of these equations in the range $0\le \phi_a < 1$.
We immediately notice that when $\phi_a = \phi_b$ for $a\neq b$ the configurations do not contribute to the partition function.
All the $\phi_a$ are contained in the range, even if we interchange the all solutions $\phi_a$.
So we can set $\phi_1 < \phi_2 < \cdots < \phi_k$ and
a factor $k!$ in the partition function cancels out.
The number of piecewise independent solutions of the equations 
in the range of $0\le \phi_a < 1$ is 
\begin{eqnarray}
\frac{(k + N -1)!}{(N-1)!~ k!}.
\label{number primary SU(N)}
\end{eqnarray}
This number coincides with the number of primary fields of the $SU(N)_k$ WZW model.
The each solution (\ref{Solution}) is in one-to-one correspondence with the primary fields of the $SU(N)_k$ WZW model
or the highest weights of the integrable representation in the affine Lie algebra $\widehat{\mathfrak{su}}_k(N)$.
When the set of this solutions is denoted by $\{\mathrm{Sol}\}$, the partition function is
\begin{eqnarray}
Z_{\rm GWZW}^{U(N)}(\Sigma_h) = \alpha \beta^{1-h} \sum_{\phi_1,\cdots,\phi_k\in\{\mathrm{Sol}\}} 
 \left\{\frac{\prod_{\substack{a,b=1\\a \neq b}}^k 
(e^{2\pi i \phi_a}-e^{2\pi i\phi_b})}{\prod_{a=1}^k e^{2\pi i (k-1)\phi_a}}\right\}^{1-h}. 
\label{partition U(N)}
\end{eqnarray}
where $\alpha$ and $\beta$ are a genus independent and dependent constant, respectively.

Finally, we determine the normalization for the partition function 
of the $G/G$ gauged WZW model on $\Sigma_h$ 
which is compatible with the number of the conformal blocks in the $G$ WZW model on $\Sigma_h$.
The partition function of the $G/G$ gauged WZW model on $\Sigma_h$
also can be represented by the modular S-matrix for the character in the $G$ WZW model as follows
\begin{eqnarray}
Z_{\rm GWZW}^{G}(\Sigma_h) =
 \sum_{\cal R} ({\cal S}_{0{\cal R}}^{G})^{2-2h},
 \label{partition modular}
\end{eqnarray}
where ${\cal R}$ denotes an integrable highest weight representation in the affine Lie algebra $\hat{\mathfrak{g}}$ corresponding to a primary field in $U(N)$ WZW model
and the summation runs through all the primary fields in the $G$ WZW model.
Therefore, we determine the normalization  such that the partition function (\ref{partition U(N)}) matches with (\ref{partition modular}) in $G=U(N)$.
When $h=1$, the partition function of the $G/G$ gauged WZW model coincides with the number of the primary fields in $G$ WZW model.
The genus independent normalization factor $\alpha$ is $(N+k)/N$
 because  $\{{\rm Sol}\}$ only runs through the primary fields in the $\widehat{\mathfrak{su}}(N)_k$ WZW model
 and the number of the primary fields in the $\hat{\mathfrak{u}}(N)_k$ WZW model is 
\begin{eqnarray}
\frac{(k + N)!}{N!~ k!}.
\label{number primary U(N)}
\end{eqnarray}
The resulting partition function of the $U(N)/U(N)$ GWZW model on $\Sigma_h$ is 
\begin{eqnarray}
Z_{\rm GWZW}^{U(N)}(\Sigma_h)=\frac{N+k}{N}
 \sum_{\phi_1,\cdots,\phi_k\in\{\mathrm{Sol}\}} 
 \left\{\frac{1}{(k+N)^{k}}\frac{\prod_{\substack{a,b=1\\a \neq b}}^k 
(e^{2\pi i \phi_a}-e^{2\pi i\phi_b})}{\prod_{a=1}^k e^{2\pi i (k-1)\phi_a}}\right\}^{1-h}.
\label{partition U(N)}
\end{eqnarray}

\section{The phase model}
\label{sec:phase}

The phase model is a quantum integrable field theory on one-dimensional lattice
and a strongly correlated boson system.
We can calculate observables
such that energy eigenvalues, eigenvectors and correlation functions and so on in this model, by applying the algebraic Bethe Ansatz to this model.
This model is considered by e.g. \cite{Bogoliubov:1996}, \cite{Bogoliubov:1997}, \cite{Korff:2010}.
In this section, we review the phase model and the algebraic Bethe Ansatz for this model.

\subsection{The phase model}
In this subsection, we define the phase model.
First of all, we define the phase algebra.
The phase algebra $\Phi$ is an algebra such that operators 
$\{\hat{N},\varphi,\varphi^{\dagger}\}$ obey
\begin{eqnarray}
&&[\hat{N},\varphi] = -\varphi,\quad
[\hat{N},\varphi^{\dagger}] = \varphi^{\dagger},\quad
\varphi\varphi^{\dagger}=1.
\label{phase algebra}
\end{eqnarray}
The operators $\varphi$ and $\varphi^{\dagger}$ serve as an annihilation operator and a creation operator, respectively.
Next, we define a Fock space ${\cal F}$ for the phase algebra given by the equations (\ref{phase algebra}).
A vacuum state $|0\rangle$ is defined as $\varphi|0\rangle =0$.
The set $\{|m\rangle := (\varphi^{\dagger})^m|0\rangle ~|~ m\in \Z_{\ge 0}\}$
forms the basis of the Fock space 
by successive actions of the creation operator $\varphi^{\dagger}$ on the vacuum state $|0\rangle$.
It also holds following relations,
\begin{eqnarray}
\hat{N}|m\rangle = m|m\rangle,
\quad 
\varphi^{\dagger}|m\rangle = |m+1\rangle,
\quad
\varphi|m\rangle = | m-1\rangle.
\end{eqnarray}

The Hamiltonian of the phase model on the one-dimensional lattice with the total site number $N$ 
is given by 
\begin{eqnarray}
H = -\frac{1}{2}\sum_{i=1}^N \left(\varphi_i\varphi^{\dagger}_{i+1} + \varphi^{\dagger}_i\varphi_{i+1}\right)
\label{phase Hamiltonian}
\end{eqnarray}
where we imposed a periodic boundary condition $N+1 \equiv 1$ 
and set the lattice spacing $\Delta=1$.
The operators $\{\varphi_i,\varphi_i^{\dagger},\hat{N}_i\}_{i=1,\cdots,N}$
obey the $N$-fold tensor product $\Phi^{\otimes N}$ of the phase algebra (\ref{phase algebra}). 
$\Phi^{\otimes N}$ is defined by
\begin{eqnarray}
\varphi_i\varphi_j = \varphi_j\varphi_i,
\quad\varphi^{\dagger}_i\varphi_j^{\dagger} &=& \varphi_j^{\dagger}\varphi_i^{\dagger},
\quad \hat{N}_i \hat{N}_j=\hat{N}_j \hat{N}_i\nonumber\\
\hat{N}_i\varphi_j - \varphi_j \hat{N}_i =-\delta_{ij}\varphi_i&,&
\hat{N}_i\varphi^{\dagger}_j - \varphi_j^{\dagger} \hat{N}_i = \delta_{ij}\varphi_i^{\dagger}\nonumber\\
\varphi_i\varphi_i^{\dagger}=1&,&
\varphi_i\varphi_j^{\dagger} = \varphi_j^{\dagger}\varphi_i~~{\rm if}~i\neq j\nonumber\\
\hat{N}_i(1 - \varphi_i^{\dagger}\varphi_i) = &0& = (1-\varphi_i^{\dagger} \varphi_i)\hat{N}_i.
\label{phase algebra tensor}
\end{eqnarray}
where indices $i,j$ label the sites of the lattice.
Therefore, the Hamiltonian belongs to $\Phi^{\otimes N}$ and acts on the $N$-fold tensor products of the Fock space ${\cal F}^{\otimes N}$.
The basis of ${\cal F}^{\otimes N}$ consists of
$\{|m_1,\cdots,m_N\rangle:=|m_1\rangle\otimes\cdots\otimes|m_N\rangle ~| m_i\in \Z_{\ge 0}\}$.

To better understand this model, let us change the operators obeying the phase algebra 
to the operators $\{a_i,a^{\dagger}_i,\hat{N}_i\}_{i=1,\cdots,N}$ obeying the free boson algebra 
\begin{eqnarray}
[N_i, a_j] = -\delta_{ij} a_j,\quad
[N_i,a_j^{\dagger}] = \delta_{ij} a^{\dagger}_j,\quad
[a_i,a_j^{\dagger}] = \delta_{ij},\quad N_i = a_i^{\dagger}a_i,
\end{eqnarray}
as follow
\begin{eqnarray}
\varphi_i = 
\frac{1}{\sqrt{1+\hat{N}_i}} a_i,\quad
\varphi_i^{\dagger}
= a_i^{\dagger}
\frac{1}{\sqrt{1+\hat{N}_i}}
\label{substitution}
\end{eqnarray}
where $(1+\hat{N}_i)^{-1/2}$ is defined as formal power series.
Substituting (\ref{substitution}) into  the Hamiltonian (\ref{phase Hamiltonian}), 
we found that the Hamiltonian has infinite interaction terms in front of the hopping term.
Therefore we found that the phase model is  the strongly interacting system
and the field theory with non-local interactions on the lattice.

\subsection{Algebraic Bethe Ansatz for the phase model}
In this subsection,
we consider the algebraic Bethe Ansatz for the phase model.
We follow the convention of \cite{Korff:2010}.
The L-matrix of the phase model at a site $n~(n=1,\cdots,N)$ is defined by
\begin{eqnarray}
L_n(\mu) = \left(
\begin{array}{cc}
1&~~\mu\varphi^{\dagger}_n\\
\varphi_n&~~\mu
\end{array}
\right)
\quad
\in {\rm End}[\C^2(\mu)]\otimes \Phi,
\end{eqnarray}
where $\mu \in \C$ is a spectral parameter.
Here, the L-matrix is a matrix in an auxiliary space $\C^2$.
This L-matrix satisfies the Yang-Baxter equation
\begin{eqnarray}
R(\mu,\nu) (L(\mu)\otimes L(\nu)) =  (L(\nu)\otimes L(\mu)) R(\mu,\nu),
\label{YB}
\end{eqnarray}
with the R-matrix 
\begin{eqnarray}
R(\mu,\nu) = \left(
\begin{array}{cccc}
\frac{\mu}{\mu-\nu}&0&0&0\\
0&\frac{\nu}{\mu-\nu}&1&0\\
0&0&\frac{\mu}{\mu-\nu}&0\\
0&0&0&\frac{\mu}{\mu-\nu}
\end{array}
\right)
\quad
\in {\rm End}[\C^2(\mu)\otimes\C^2(\nu)].\label{R matrix}
\end{eqnarray}
The monodromy matrix is defined by
\begin{eqnarray}
T(\mu) = L_N(\mu)L_{N-1}(\mu)\cdots L_1(\mu)
=\left(
  \begin{array}{cc}
    A(\mu)   & B(\mu)   \\
     C(\mu)   &  D(\mu)    \\
  \end{array}
\right).
\label{monodoromy matrix}
\end{eqnarray}
From the Yang-Baxter equation (\ref{YB}), the monodromy matrix satisfies a following relation: 
\begin{eqnarray}
R(\mu,\nu) (T(\mu)\otimes T(\nu)) =  (T(\nu)\otimes T(\mu)) R(\mu,\nu).
\label{TTR relation}
\end{eqnarray}
From this relation, we can derive commutation relations of the monodromy matrix elements,
$A(\mu),B(\mu),C(\mu),D(\mu)$:
\begin{eqnarray}
&&B(\mu)B(\nu) = B(\nu)B(\mu)\\
&&C(\mu)C(\nu) = C(\nu)C(\mu)\\
&&(\mu-\nu)A(\mu)B(\nu) = -\nu B(\nu)A(\mu) + \nu B(\mu)A(\nu) \label{commutationAB}\\
&&(\mu-\nu)D(\mu)B(\nu) = \mu B(\nu)D(\mu) - \nu B(\mu)D(\nu)\label{commutationDB}\\
&&C(\mu)B(\nu) 
= \frac{\nu}{\mu-\nu}(A(\nu)D(\mu) - A(\mu)D(\nu))\label{commutationCB2}
\end{eqnarray}
Taking trace with respect to the auxiliary space, the monodromy matrix becomes the transfer matrix
\begin{eqnarray}
\tau(\mu) = \mathrm{tr}T(\mu) = A(\mu) +  D(\mu).
\end{eqnarray}
The vacuum state $|0\rangle$ and its dual vacuum state $\langle 0|$ satisfy 
\begin{eqnarray}
C(\mu)|0\rangle=0\quad{\rm and}\quad\langle 0|B(\mu) = 0,
\end{eqnarray}
because $C(\mu)$ is the creation operator and $B(\mu)$ is the annihilation operator.
$a(\mu)=1$ and $d(\mu) = {\mu}^N$ are the eigenvalues of operators $A(\mu)$ and $D(\mu)$ on the vacuum state $|0\rangle$, respectively.

Eigenstates of the transfer matrix can be constructed 
by repeated actions of the operators $B(\lambda)$ on the vacuum state $|0\rangle$,
that is to say, a state $\prod_{j=1}^k B(\lambda_j^{-1})|0\rangle$ is the eigenstate of the transfer matrix
\begin{eqnarray}
\tau(\mu)\prod_{a=1}^k B(\lambda_a^{-1})|0\rangle =\Lambda(\mu,\{\lambda\})\prod_{a=1}^k B(\lambda_a^{-1})|0\rangle
\end{eqnarray}
where
\begin{eqnarray}
\Lambda(\mu,\{\lambda\})=\prod_{a=1}^k \frac{1}{1-\lambda_a \mu} +\mu^N \prod_{a=1}^k \frac{\mu\lambda_a}{\mu\lambda_a-1},
\label{transfer matrix eigen value}
\end{eqnarray}
if the spectrum parameters $\{\lambda_j\}_{j=1,\cdots,k}$ satisfy Bethe Ansatz equations
\begin{eqnarray}
(-1)^{k-1}\cdot\lambda_a^{(k+N)}\cdot\prod_{b=1}^k\lambda_b^{-1}  = 1.
\label{Bethe eq}
\end{eqnarray}

Finally, we consider a Bethe norm  which
is a norm of the Bethe vector
$|\psi(\{\lambda\}_k)\rangle = \prod_{a=1}^k B(\lambda_a^{-1})|0\rangle$
and its dual vector 
$\langle \psi(\{\lambda\}_k)| = \langle 0|\prod_{a=1}^k C(\lambda_a^{-1})$.
Here, $\{\lambda_j\}_{j=1,\cdots,k}$ satisfy the Bethe Ansatz equations.
The Bethe norm in the phase model is given by
\begin{eqnarray}
\langle \psi(\{\lambda\}_k)|\psi(\{\lambda\}_k)\rangle
= \frac{\prod_{b=1}^k \lambda_b^{k-1}}{\prod_{\substack{a,b=1\\a\neq b}}^k (\lambda_a - \lambda_b)}
N(N+k)^{k-1}.
\label{Bethe norm}
\end{eqnarray}
See \cite{Bogoliubov:1997},\cite{Korepin:book} for the derivation.
In the next section, we will clarify  connections
between the $U(N)/U(N)$ gauged WZW model and the phase model.

\section{The $U(N)/U(N)$ gauged WZW model and the phase model}
\label{sec:GWZW-phase}

In this section, we clarify connections between the $U(N)/U(N)$ gauged WZW model and the phase model.
First of all we have to identify parameters of both theories.
We identify the level $k$ and the rank $N$ of the gauge group $U(N)$ in the $U(N)/U(N)$ gauged WZW model
with the total particle number $k$ and the total site number $N$ in the phase model, respectively.
Under these parameter identifications, we can show that the constraints (\ref{constraint}) coincide with the Bethe Ansatz equations (\ref{Bethe eq}). 
Taking the parameterization of the Bethe roots as $\lambda_a = e^{2\pi i\phi_a}$,
the logarithm form of the Bethe Ansatz equations becomes
\begin{eqnarray}
(N+k) \phi_a - \sum_{b=1}^k \phi_b + \frac{k-1}{2}= m_a
\label{log Bethe equation}
\end{eqnarray}
where $m_a\in \Z$ implies branches of the logarithm.
Once we identify the constant field $\phi_a$ in the $U(N)/U(N)$ gauged WZW model with the Bethe roots $\phi_a$ in the phase model,
we found that these equations coincide with the localization configurations (\ref{constraint}) 
in the $U(N)/U(N)$ gauged WZW model.

Next, let us consider solutions of the Bethe Ansatz equations.
The solutions  are (\ref{Solution}) because the Bethe Ansatz equations are equal to the localized configurations (\ref{constraint}).
Then we can show that piecewise independent solutions of the Bethe Ansatz equations coincide with the solutions to be included in the range of $0\le \phi_a <1$ 
and to satisfy the condition $0<\phi_1<\phi_2<\cdots<\phi_k$ in the $U(N)/U(N)$ gauged WZW model.
Thus, we found that this solutions of the Bethe Ansatz equations  coincide with $\{\text{Sol}\}$.
The solutions (\ref{Solution}) also imply the completeness of the state in the phase model 
because the number of the solutions is $(N+k-1)!/(N-1)!k!$.

Since the Bethe norm in the phase model (\ref{Bethe norm}) becomes
\begin{eqnarray}
\langle \psi(\{e^{2\pi i \phi}\}_k)|\psi(\{e^{2\pi i \phi}\}_k)\rangle
=  \frac{\prod_{a=1}^k e^{2\pi i (k-1)\phi_a}}{\prod_{\substack{a,b=1\\a \neq b}}^k 
(e^{2\pi i \phi_a}-e^{2\pi i\phi_b})}(k+N)^{k-1}N
\label{Bethe norm2}
\end{eqnarray}
under taking the parameterization of the Bethe roots as $\lambda_a = e^{2\pi i\phi_a}$,
the partition function (\ref{partition U(N)}) of the $U(N)/U(N)$ gauged WZW model can be represented as
\begin{eqnarray}
Z_{\rm GWZW}^{U(N)}(\Sigma_h) = \left(\frac{N+k}{N}\right)^{h}
 \sum_{\phi_1,\cdots,\phi_k\in\{\mathrm{Sol}\}} 
\langle \psi(\{e^{2\pi i \phi}\}_k)|\psi(\{e^{2\pi i \phi}\}_k)\rangle^{h-1}.
\label{partition U(N) phase}
\end{eqnarray}

Why the partition function of the $U(N)/U(N)$ gauged WZW model can be represented by the Bethe norm in the phase model?
To understand this, we recall that the partition function is represented by using the modular S-matrix 
(\ref{partition modular}).
Thus, we can expect that there is a relation between the modular S-matrix in $U(N)/U(N)$ gauged WZW model and the Bethe norm in the phase model.
Actually, Korff and Stroppel constructed the Verlinde algebra  in the $SU(N)$ WZW model on the sphere from a viewpoint of the phase model and showed that the modular S-matrix in $SU(N)$ WZW model coincides with the Bethe norm in the phase model \cite{Korff:2010}.
So, let us derive the partition function of the $SU(N)/SU(N)$  gauged WZW model from the one of the $U(N)/U(N)$ case.
There are two differences between these partition functions. 
Firstly, the modular S-matrices in each model are related to 
\begin{eqnarray}
{\cal S}_{0{\cal R}}^{\hat{\mathfrak{u}}(N)} = \sqrt{\frac{N}{N+k}}{\cal S}^{\widehat{\mathfrak{su}}(N)}_{0R}.
\end{eqnarray}
where ${\cal R}$ and $R$ denote the primary field in the $U(N)$ and the $SU(N)$ WZW model, respectively \cite{Naculich:2007nc}.
Secondary, a range which the summation runs through is different because the number of the each primary field is different.
Taking account these two differences, we find that the partition function of the $SU(N)/SU(N)$ gauged WZW model is
\begin{eqnarray}
Z_{\rm GWZW}^{SU(N)}(\Sigma_h)= \sum_{\phi_1,\cdots,\phi_k\in\{\mathrm{Sol}\}} 
 \left\{\frac{1}{(k+N)^{k-1}N}\frac{\prod_{\substack{a,b=1\\a \neq b}}^k 
(e^{2\pi i \phi_a}-e^{2\pi i\phi_b})}{\prod_{a=1}^k e^{2\pi i (k-1)\phi_a}}\right\}^{1-h} 
\label{SU(N) partition function}
\end{eqnarray}
and can be represented by the summation of the Bethe norm with respect to  all the eigenstates of the transfer matrix in the phase model;
\begin{eqnarray}
Z_{\rm GWZW}^{SU(N)}(\Sigma_h) =
 \sum_{\phi_1,\cdots,\phi_k\in\{\mathrm{Sol}\}} 
\langle \psi(\{e^{2\pi i \phi}\}_k)|\psi(\{e^{2\pi i \phi}\}_k)\rangle^{h-1}.
\label{partition SU(N) phase}
\end{eqnarray}
This shows that the modular S-matrix of the $SU(N)$ WZW model coincides with the Bethe norm;
\begin{eqnarray}
{\cal S}_{0R}^{SU(N)} =  \langle \psi(\{e^{2\pi i \phi}\}_k)|\psi(\{e^{2\pi i \phi}\}_k)\rangle.
\end{eqnarray}
This is considered as a reason why the partition function of  the $U(N)/U(N)$ gauged WZW model can be represented by the Bethe norm in the phase model.
Therefore, we found that the Gauge/Bethe correspondence between the $G/G$ gauged WZW model and the phase model is also considered as the gauged WZW model realization of \cite{Korff:2010}.

Finally, we comment relations between the CS theory and the phase model.
The partition function of the $G/G$ gauged WZW model coincides with the partition function of the CS theory with the gauge group $G$ on $S^1 \times \Sigma_h$ \cite{Blau:1993tv}.
We can apply equivariant localization methods to the CS theory in a similar way with $G/G$ gauged WZW model.
Thus in the CS theory with the gauge group $U(N)$ on $S^1 \times \Sigma_h$ , 
the localization configurations coincide with the Bethe Ansatz equations and the partition function is represented by the Bethe norm in the phase model:
\begin{eqnarray}
Z_{\text{CS}}^{U(N)}(S^1\times \Sigma_h)
=\left(\frac{N+k}{N}\right)^{h}
 \sum_{\phi_1,\cdots,\phi_k\in\{\mathrm{Sol}\}} 
\langle \psi(\{e^{2\pi i \phi}\}_k)|\psi(\{e^{2\pi i \phi}\}_k)\rangle^{h-1}.
\label{CS partition U(N) phase}
\end{eqnarray}
Further, when the gauge group is $SU(N)$, the partition function of the CS theory is
\begin{eqnarray}
Z_{\text{CS}}^{SU(N)}(S^1\times \Sigma_h)
= \sum_{\phi_1,\cdots,\phi_k\in\{\mathrm{Sol}\}} 
\langle \psi(\{e^{2\pi i \phi}\}_k)|\psi(\{e^{2\pi i \phi}\}_k)\rangle^{h-1}.
\label{CS partition SU(N) phase}
\end{eqnarray}

We have shown the Gauge/Bethe correspondence between CS theory on $S^1 \times \Sigma_h$ and the phase model. The equivariant localization for the CS theory on wider class manifolds (Seifert manifolds) is derived in \cite{Beasley:2005vf}, \cite{Blau:2006gh} and \cite{Kallen:2011ny}, see also \cite{Ohta:2012ev} for generalization to the Chern-Simons-Matter theories. To describe the partition function of the CS theory on these manifolds,  not only modular S-matrix but also modular T-matrix is needed. It would be interesting to consider meaning of T-matrix in the phase model side.

\section{Conclusion}
\label{sec:summary}

In this paper, we have studied the relation between the $U(N)/U(N)$ gauged WZW model and the algebraic Bethe Ansatz for the phase model.
We found that the localization configurations (\ref{constraint}) coincide with the Bethe Ansatz equations (\ref{log Bethe equation}), once the diagonal group elements, the level and the rank of the gauge group $U(N)$  in the $U(N)/U(N)$ gauged WZW model 
are  identified with the Bethe roots,  the total site number and the total particle number in the phase model, respectively.
We also showed that the partition function of the $U(N)/U(N)$ and the $SU(N)/SU(N)$ gauged WZW model is represented as the summation of the Bethe norm with respect to the all eigenstates of the transfer matrix in the phase model.
This is because the modular S-matrix in the $SU(N)$ WZW model coincides with the Bethe norm.
This is also considered as the gauged WZW model realization involving a generalization  to a higher genus case of \cite{Korff:2010}.
We further found that the partition function of the CS theory on $S^1\times \Sigma_h$ is also related to norms of  Hamiltonian eigenstates for the phase
model.
These relations are summarized in the table 1.

\begin{table}[h]
 \begin{center}
  \begin{tabular}{|c|c|}
    \hline
  the phase model    &  the U(N)/U(N) GWZW model/the U(N) CS theory\\
   \hline \hline
 Bethe root    & diagonal group element/holonomy along $S^1$ direction  \\
    \hline
 Bethe Ansatz equation & configuration of (\ref{constraint})  \\
    \hline
 Total site number     & rank of the gauge group $U(N)$ \\
    \hline
 Total particle number  &  level  \\
    \hline
 Bethe norm   & modular S-matrix ${\cal S}_{0\hat{\mu}}$\\
 \hline
  \end{tabular}
\end{center}
\caption{Dictionary in the Gauge/Bethe correspondence}
 \label{tb:correspondence}
\end{table}

We comment on some directions for future works. 
We have considered the case of the gauge group $U(N)$ or $SU(N)$ in the $G/G$ gauged WZW model.
It is interesting to consider a generalization to an arbitrary semi-simple gauge group $G$. 

Equivariant localization formula itself is not restricted to the topological Yang-Mills theory \cite{Witten:1992xu} nor the $G/G$ gauged WZW model.
One  can also  couple  additional matter fields to the topological Yang-Mills theory and  
the $G/G$ gauged WZW model, and perform the localization. Then, these systems will correspond to other integrable systems.
For an example, the localization configuration for the $G/G$ gauged WZW model coupled to a one-form Higgs field leads to the 
Bethe Ansatz equations for an infinite spin limit for higher spin XXZ model \cite{Gerasimov:2006zt}. 
The relation between the partition function of the $G/G$ gauged WZW + Higgs system and the norm of Bethe vector for XXZ model will be studied in detail \cite{Okuda}. 

Another example is following: Nekrasov and Shatashvili found that a twisted superpotential in $\mathcal{N}=(2,2)$ supersymmetric gauge theory coincides  with
a Yang-Yang function for corresponding integrable system \cite{Nekrasov:2009ui}, \cite{Nekrasov:2009uh}. 
When we consider the topologically twisted supersymmetric gauge theory on a compact Riemann surface with same matter content of \cite{Nekrasov:2009ui}, \cite{Nekrasov:2009uh}, the localization configuration will lead to 
the same Bethe Ansatz equation which arises from the saddle point equations for  the effective twisted superpotential of $\mathcal{N}=(2,2)$ theory. 
For the localization of ${\cal N}=(2,2)$ supersymmetric gauge theory on $S^2$, see \cite{Benini:2012ui}, \cite{Doroud:2012xw}.
It  would be interesting to study localization for such  topological twisted gauge theories and correspondence with integrable systems.

\subsection*{Acknowledgments}
We are grateful to Tetsuo~Deguchi, Tohru~Eguchi, Kazutoshi~Ohta and   Katsuyuki~Sugiyama  for useful discussions
and comments. Y.~Y is also grateful to  Satoshi~Yamaguchi for discussions and his hospitality during visit to Osaka University.

%\newpage

\end{document}